\def \s{\begin{equation}\addtocounter{equation}{-1}}
\def \e{\end{equation}}
\def \labep{\addtocounter{equation}{+1}\label}
\documentstyle[11pt,aaspp4,flushrt]{article}

\begin{document}
\def\huh{\hbox{\vrule width 2pt height 8pt depth 2pt}}
\def\infootbegplain{}
\def\infootendplain{}
\def\rfn{\footnote}                          
\def\qfn{\def\acomment}                      
\def\pfn{\def\acomment}                      
\def\ifn{\def\acomment}                      
\def\smf{\tiny \parindent=0.0truecm \parskip=0.0truecm \baselineskip=0.0truecm}
\def\infoot{\input}  
\def\blacksquare{\hbox{\vrule width 4pt height 4pt depth 0pt}}
\def\square{\hbox{\vrule\vbox{\hrule\phantom{o}\hrule}\vrule}}
\def\fig#1#2{\begin{figure}
 \centering
 \plotone{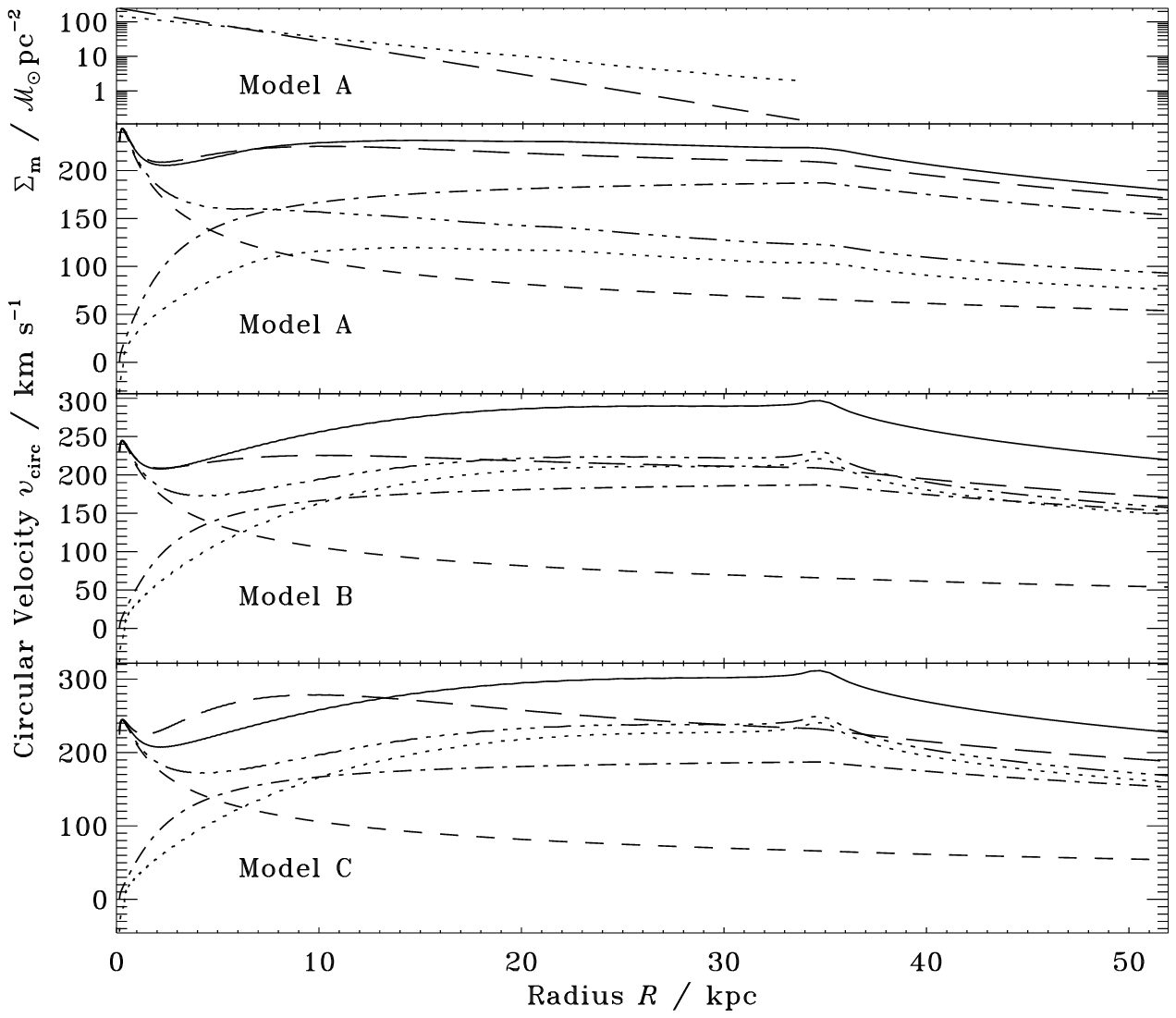}
 \figcaption{\setlength{\baselineskip}{.2in}#2}
 \label{#1}
 \end{figure}}
\def\Fig#1#2{\fig{#1}{#2}Figure 1}
\def\inputfigurecaptionshere{}
\def\inputfigureplotshere{}
%
%
%
%
%
%
%
\infootbegplain
\title{Mass Function Gradients and the Need for Dark Matter}
\affil{Accepted for publication in {\it The Astrophysical Journal Letters}}
\author{Jason A.~Taylor\altaffilmark{1,2}}
\affil{Department of Physics, United States Naval Academy, Annapolis, MD
21402-5026, USA; E-mail (Internet): taylor@milkyway.gsfc.nasa.gov}
\altaffiltext{1}{Postal address: NASA/GSFC, Laboratory for High Energy
Astrophysics, Code 661, Greenbelt  MD~~20771, USA}
\altaffiltext{2}{Department of Physics, University of Maryland, College Park,
MD~~20742, USA}
\par
\noindent \begin{abstract}There
is both theoretical and empirical evidence that the initial mass function
(IMF) may be a function of the local star formation conditions.  In
particular, the IMF is predicted to flatten with increasing local luminosity
density $\rho _{l}$, with the formation of massive stars being preferentially
enhanced
in brighter regions.  In R136, the bright stellar cluster in 30 Doradus, the
IMF gradient is $\partial \Gamma /\partial \log \rho _{{\scriptsize\rm
l}}=0.28\pm 0.06,$ where $\Gamma $ is the slope of the IMF.  If such
IMF gradients are indeed general features of
galaxies, this implies that several previous astrophysical measurements, such
as the surface mass densities of spirals (obtained assuming constant mass to
light ratios), were plagued by substantial systematic errors.  In this Letter,
calculations which account for possible IMF gradients are presented of surface
densities of spiral galaxies.  Compared to previous estimates,
the mass surface densities corrected for IMF gradients are higher in the outer
regions of the disks.  For a model based on the Milky Way but with an IMF
scaled according to R136, the rotation curve without the traditional dark halo
component falls with Galactocentric radius, though slower than it would
without IMF gradients.  For a second model of the Milky Way in which the IMF
gradient is increased to 0.42, the rotation curve is approximately flat in the
outer disk, with a rotational velocity below $\simeq 220$ km
s$^{{\scriptsize\rm -1}}$ only before the
traditional dark halo component is added.  For a third model in which
substantial arm/interarm density contrasts are additionally assumed, the solar
vicinity mass density drops to $0.10 M_{\odot }$pc$^{-3}$, which is consistent
with
observations. These results, if generalizable to other galaxies, not only
call into question the assertion that dark matter halos are compatible with
the flat rotation curves of spiral galaxies, but also may clarify our
understanding of a wide variety of other astrophysical phenomena such as the
G-dwarf problem, metallicity gradients, and the Tully-Fisher relation.
\end{abstract}
{}\setcounter{section}{1}\addtocounter{section}{-
1}\section{INTRODUCTION}\def\figdirprefix{/home/theory1/taylor/stellarmodels/}
In a
recent paper, Padoan, Nordlund, \& Jones (1997) claimed on theoretical grounds
that the initial mass function (IMF) should be a function of the local
temperature ${\it T}$ of the original molecular clouds.  Padoan et al. (1997)
argued that dense star forming regions, such as those in starburst galaxies,
should be warmer than sparser star forming regions.  In fact, if the
temperature dependence of the clouds is not drastically different from that of
a blackbody, then $T\propto \rho ^{1/4}_{{\scriptsize\rm l}}$, where $\rho
_{{\scriptsize\rm l}}$ is the local mean luminosity once star
formation has already started.  Padoan et al. claimed that starburst regions
should therefore have a flatter IMF and be more ``top heavy."  Similar
reasoning would imply that the IMF in cooler regions of galaxies should favor
low mass star formation and be steeper.
\par
In support of their star formation model, Padoan et al. (1997) noted that
for $T\gtrsim 60$ K, their models predict a top heavy IMF similar to that
found in the
center of R136~(Malumuth \& Heap 1994, Brandl et al. 1996), the bright stellar
cluster in 30 Doradus.  Due to its proximity
and the fact that it is the most massive H {\scriptsize\rm II} region in the
Local Group, 30
Doradus is perhaps the best star formation ``laboratory" accessible to us.
However, the relaxation time in R136 may be
less than its age (Campbell et al. 1992), so dynamic friction may also
contribute toward the R136 present-day mass function gradient.
\par
Fortunately, many other avenues of testing Padoan et al.'s model exist.
The O-star catalog of Garmany, Conti, \& Chiosi (1982) shows a flattening of
the IMF slope toward the Galactic center (cf Humphreys \& McElroy 1984).  This
data supports Padoan et al.'s model since higher surface brightness regions
would, on the average, yield higher temperatures and flatter IMFs.  Models
which
attempt to explain correlations between local surface brightness, color, line
ratios, metallicity, and the star formation rate have assumed
luminosity-dependent IMFs (e.g., Edmunds \& Phillipps 1989; Phillipps,
Edmunds,
\& Davies 1990).  Several evolutionary models of inner regions of starburst
galaxies assume low mass cutoffs or top heavy IMFs (e.g., Rieke et al. 1980;
Augarde \& Lequeux 1985; Doane \& Mathews 1993; Doyon, Joseph, \& Wright
1994).
Finally, independent theoretical arguments supporting
IMF gradients range from models which are consistent with the simple form of
the Jeans expression for the typical stellar mass in solar units of
$<m>\propto T^{3/2}~($e.g.; Larson
1982; Bodenheimer, Tohline, \& Black 1980) to much more complicated models,
such as the outflow-regulated model of Adams \& Fatuzzo (1996), which predicts
$<m>\propto T^{a}$, where $1\le a\le 3/2.$
\par
If IMFs are actually a function of $\rho _{{\scriptsize\rm l}}$ or $T$, there
would several important
astrophysical consequences.  For instance, there would be a
position-dependence in the mean mass to light ratio.  This is due to the
strong dependence of the mass to light ratio upon the IMF.  In R136, this
makes the mass density function $\rho _{{\scriptsize\rm m}}$ much different
from $\rho _{{\scriptsize\rm l}}~($Malumuth \&
Heap 1994, Brandl et al. 1996) and complicates
estimates of the total mass.  Padoan et al.'s
results indicate that similar effects might occur in spiral galaxies.  If the
luminosity of a star is taken as $L\simeq L_{\odot }m^{y}$, where $y\simeq
3.5,$ the Jeans expression
above would suggest the crude relation $<m>\propto \rho
^{3/8}_{{\scriptsize\rm l}}$
and yield $\rho _{{\scriptsize\rm m}}\propto \rho ^{1+3(1-
y)/8}_{{\scriptsize\rm l}}\simeq \rho ^{0.06}_{{\scriptsize\rm l}}$.
Unfortunately, previous
works
have assumed that IMFs are independent of time and position with,
specifically, $\rho _{{\scriptsize\rm m}}\propto \rho ^{1.0}_{{\scriptsize\rm
l}}$ throughout a given spiral galaxy (e.g., van Albada et
al. 1985).  In this {\it Letter}, surface mass densities of spiral galaxies
are
computed, for the first time, by explicitly accounting for the possible types
of IMF gradients that might exist if theories like those of Padoan et al. are
correct.
\par
\noindent {}\setcounter{section}{2}\addtocounter{section}{-1}\section{AN
EMPIRICAL ESTIMATE OF THE R136 IMF GRADIENT}
\par
Since position-dependent measurements in R136
of both $\rho _{{\scriptsize\rm l}}$ and the IMF slope $\Gamma $ (where ${\it
dN}/{\it dM\propto m}^{\Gamma -1}$ is the number of stars per
unit mass in solar units) have already been made, computing the dependence of
the R136 IMF upon the local luminosity is straightforward.  Doing this will
provide a useful
starting point in obtaining a crude yet quantitative estimate of the possible
types of IMF gradients that might generally exist in all galaxies including
the Milky Way.
\par
Table 1 summarizes Brandl et al.'s (1996) results for the IMF based upon
high resolution 5-color photometry of the stars in R136 estimated to be
between 2.5 and 3.5 Myrs years old.  The right-most entry
of Table 1 shows the results of
performing the coordinate transformation between ${\it R}$ and $\rho
_{{\scriptsize\rm l}}$ using Figure 15 of
Hunter et al. (1996).  Though Brandl et al. (1996) did not make explicit
measurements of the upper and lower stellar mass cutoffs $m_{{\scriptsize\rm
l}}$ and $m_{{\scriptsize\rm u}}$ to the
power-law approximation of the IMF, Table 1 includes estimates of their
dependences upon the local surface brightness.  The lower mass limits were
obtained from the peaks of Brandl et al.'s mass functions, while the upper
limits were taken from the highest masses observed per radius bin. Both
$\log _{10}(m_{{\scriptsize\rm l}})$ and $\log _{10}(m_{{\scriptsize\rm u}})$
are found to decrease by $\simeq 0.2$ with each successive
increase in radius.  Brandl et al. (1996) performed completeness corrections,
so the depletion of low mass stars in all but the outer regions of R136 is
presumably real.  The results of performing a linear fit of the IMF parameters
of R136 to $\log _{10}(\rho _{{\scriptsize\rm l}})$ are shown in
columns 2-7 of Table 2 as Model A. Uncertainties of parameters calculated from
more than two radius bins are shown in parenthesis.
\par
\noindent {}\setcounter{section}{3}\addtocounter{section}{-
1}\section{DYNAMICAL PROPERTIES OF SPIRAL GALAXIES WITH IMF GRADIENTS}
\par
The IMF gradient of Model A implies a surface mass density that is
different from what would be obtained were the mass to light ratio
constant.  The surface mass density for Model A, if scaled according to the
surface luminosity function suspected for the Galaxy, is shown in the top
panel of Figure {}1.  The disk scale length $R_{0}=4.5$
kpc and solar Galactocentric radius $R_{\odot }=7.8$ kpc were taken from
Kuijken \&
Gilmore's (1989a) model of the Galaxy.  For simplicity, $\rho
_{{\scriptsize\rm l}}$ at a given radius
was assumed to be constant throughout a disk thickness of
575 pc.  The surface brightness was normalized to be $22.5~L_{\odot }$pc$^{-
2}$ at $R=R_{\odot }$,
which results in $\rho _{{\scriptsize\rm l\odot }}\equiv \rho
_{{\scriptsize\rm l}}(R_{\odot })=0.037~L_{\odot }$pc$^{-3}$.  For each
radius bin, the IMF was obtained from $\rho _{{\scriptsize\rm l}}$ and the
coefficients shown in Table 2
for Model A.  This IMF was converted to present
day mass and luminosity functions by assuming (purely for simplicity) a
constant star formation rate for the past $1.0\times 10^{10}$ yrs.  The main
sequence
lifetime-luminosity-mass relationships used to obtain the mass to light ratio
as a function of the IMF were obtained from logarithmic-linear interpolation
of $m\ge 0.8$ models published by Schaller et al. (1992) for $z=0.02,$
overshooting
of the $m\ge 1.5$ stars, and standard mass loss rates.  For $m<0.8,
L|_{m=0.25}=7.8\times 10^{-4}L_{\odot },~L|_{m=0.08}=6.55\times 10^{-
9}L_{\odot }$,~and $L|_{m\le 0.07}=5.0\times 10^{-12}L_{\odot }$ were
assumed.
\par
\medskip
The surface densities both of Model A and of the constant mass to light
ratio model fall off exponentially with increasing radius.  The effective
scale length of Model A is $\simeq 7.5$ kpc, which is $\simeq 1.7$ times
larger than that of
the surface brightness function.  This increase in the scale length is a
result of the fraction of low mass stars (and the mass to light ratio)
increasing with radius.
\par
\medskip
From the surface density, other dynamical properties of the galaxy can
also be calculated.  The circular velocity (i.e., the rotation curve)
corresponding to the surface density of Model A is shown in the second panel
of \Fig{v_circstriplet}{{\it Top panel}: the surface density of a spiral
galaxy
similar to the Milky Way but with the IMF of Model A.  The dotted line is
the surface density assuming that all stars lie on the main-sequence.
The dashed line is the surface density if the {\it V} band mass to light ratio
were
constant at $\gamma _{{\scriptsize\rm V}}=2.0M_{\odot }/L_{\odot }$.  {\it
Lower three panels}: circular velocities of Models A
(upper-middle), B (lower-middle), and C (bottom). The circular velocities of
Model A correspond to the surface density function shown in the top panel.
For each model, the solid curve accounts for all components of mass, the
dot-dash curve accounts for just the halo, dotted curve accounts for just the
disk, the short-dashed curve accounts for just the bulge and spheroid stars,
the dash-dotted curve accounts for everything except the halo, and the
long-dashed curve assumes that the mass to light ratio is constant at
$\gamma _{{\scriptsize\rm V}}=2.0M_{\odot }/L_{\odot }$.}.  The parameters for
the bulge, spheroid, and halo were taken from
Table 1 and
Figure 5 of Kuijken \& Gilmore's (1989a) model of the Galaxy.  To avoid a
divergent and unphysical total mass, the additional assumption that all
components of the Galaxy terminate at an arbitrarily-selected maximum radius
of 35.0 kpc was also made.  For Model A, this results in a total mass of the
halo, bulge/spheroid, and disk of, respectively, $2.8\times 10^{11} M_{\odot
}, 3.5\times 10^{10} M_{\odot }$, and
$6.3\times 10^{10} M_{\odot }$.  In comparison, the integrated disk mass of
the $\gamma _{{\scriptsize\rm V}}=2.0M_{\odot }/L_{\odot }$ model
is only $3.2\times 10^{10} M_{\odot }$ and increases much faster with radius.
For simplicity,
the surface mass density of stellar remnants and gas was assumed throughout
the disk to be 1/3 that of the stars.  Because the halo dominates the mass
distribution, the circular velocity curve (solid line) is nearly flat.
Without the halo, the circular velocity curve falls from 185
km s$^{-1}$ at $R=2.0$ kpc to 124 km s$^{-1}$ at $R=34$ kpc. Though the
surface
density of Model A corrected for IMF gradients is different from that
previously obtained for spiral galaxies, Figure {}1 shows
that
the change is not enough to dramatically affect the dynamical properties of
the disk, such as the circular velocity curve.
\par
\medskip
The IMF of Model A is very
negative at all radii, with $\Gamma |_{R=1.0\hbox{~{\scriptsize\rm kpc}}}=-
3.2, \Gamma |_{R=R_{\odot }}=-3.4,$ and
$\Gamma |_{R=15.0\hbox{~{\scriptsize\rm kpc}}}=-3.6.$  This occurs even though
R136$'s$ spatially-averaged
IMF is typical and its IMF gradient is small only because it has a luminosity
density that is $\sim 10^{4}-10^{8}$ times higher than typical regions of
spiral galaxies.
For
comparison, Miller \& Scalo
(1979) obtained much higher values of $\Gamma =-0.4, -1.5,$ and -2.3 for,
respectively, $0.1<m<1.0, 1.0<m<10,$ and $m>10.$  This suggests that R136's
IMF is
correlated with $\rho _{{\scriptsize\rm l}}$ in a somewhat different way than
the correlation that might
exist in the Milky Way.
\par
\medskip
In retrospect, this should not be surprising because R136 is much
different than a spiral galaxy.  The bright, early-type stars in spiral
galaxies are generally confined to relatively narrow galactocentric radii
near that of their initial birth sites.
In contrast, stars in elliptical galaxies similar to R136 undergo substantial
mixing due to their highly eccentric orbits.  Therefore, the form of the IMF
in spirals could be different than that in R136.
\par
\medskip
For this reason, other models were also considered.  Model B was
constructed in order to help answer the question of just how necessary the
dark halo is for circular velocity curves to be flat.  The IMF gradient
$\Gamma _{1}$
was adjusted to minimize the curvature of the outer circular velocity curve,
while $\Gamma _{0}$ was adjusted such that the rotation velocity was $\simeq
220$ km s$^{-1}$.
For simplicity, $m_{{\scriptsize\rm l}}$ and $m_{{\scriptsize\rm u}}$ were
fixed.  The middle plot of Figure {}1 shows that the
circular velocity curve of Model B is surprisingly flat throughout most of the
outer regions of the disk before the halo component is included.  The total
disk mass for Model B is $2.4\times 10^{11}M_{\odot }$, with $<\gamma
_{{\scriptsize\rm v}}>_{\hbox{{\scriptsize\rm disk}}}{\scriptsize\rm
=}14.5M_{\odot }/L_{\odot }$, which is 7.3
times
larger than the $\gamma _{{\scriptsize\rm V}}=2.0M_{\odot }/L_{\odot }$ model.
The value of $\Gamma _{1}$ for Model B is 0.42.  This
is
$50\%$ higher than the IMF gradient in R136.  The change within the Milky Way
of
$\Gamma $ measured by Garmany et al. (1982) between the inner and outer
semicircular
regions of radius 2.5 kpc surrounding the Sun was -0.8, which for a disk scale
length of $R_{0}=4.5$ kpc corresponds to $\Gamma _{1}=0.8\times 3\pi
R_{0}/(8$log{\it e}$\times 2.5$~kpc)=3.9.  This is
much higher than the value in Model B.  Thus the IMF gradient of Model B is
well within empirical upper limits.
\par
\medskip
However, there are at least five potential problems with the
halo-less
form of Model B: ${\bf 1}{\hbox{\bf)}}$ In the adopted solar vicinity
$(R_{\odot }=7.8$ kpc), the surface
density is $179M_{\odot }$pc$^{-2}$.  This is an unacceptable 15
standard deviations higher than the local value of $46\pm 9~M_{\odot }$pc$^{-
2}$ measured by
Kuijken \& Gilmore (1989b).  The corresponding mass to light ratio is
$7.9M_{\odot }/L_{\odot }$.
This is $60\%$ higher than the local value adopted in standard texts such as
Binney \& Tremaine (1987).  Similarly, the IMF slope at this radius is $\Gamma
=-1.7.$
Also, for the low mass $(m\lesssim 0.5)$ stars, this value is incompatible
with Miller \&
Scalo's (1979) result of $\Gamma =-0.4.~~{\bf 2}{\hbox{\bf)}}$ Mestelian
disks, which are similar to
Model B, are
commonly thought to be unstable to bar formation.  The Toomre instability
parameter $Q$ is $\sigma _{R}\kappa /(2.9G\Sigma _{{\scriptsize\rm m\odot
}})$, where $\kappa \simeq 36$ km s$^{-1}$ kpc$^{-1}$ is the epicycle
frequency and $\sigma _{R}$ is the mass-weighted stellar velocity dispersion
(Toomre
1974).  Published estimates are $Q\simeq 1-3$ in the solar vicinity.  Because
        stellar
velocity dispersions are empirically observed to decrease with mass even for
stars with lifetimes greater than the age of the galaxy, estimates of $\sigma
_{R}$ are
sensitive to $m_{{\scriptsize\rm l}}$.  Wielen (1977) obtained $\sigma
_{R}=62\pm 12$ km s$^{-1}$  for $0.1\lesssim m\lesssim 0.8$~K and
M dwarfs, which implies $Q\gtrsim 1.0\pm 0.2$ for Model B.  This lower limit
is low enough
to sustain spiral arm structure which numerical simulations show would rapidly
dissipate otherwise.  However, it is too near unity to prevent the growth of
substantial arm/interarm stellar mass density contrasts.  Though such mass
contrasts are now known to exist in normal spirals (e.g., Rix \& Zaritsky
1995,
Gonz\'alez \& Graham 1996), they are not accounted for in Model B.
Incidentally,
the halo component does not necessarily affect this instability (Sellwood
$1985).~~{\bf 3}{\hbox{\bf)}}$ The circular velocity curve at $R\lesssim 35$
kpc is not precisely flat,
but actually rises before attaining a nearly Keplerian fall off.  This is the
result of the non-spherical potential.
${\bf 4}{\hbox{\bf)}}$ The circular velocity drops below 200 km s$^{-1}$ in
the inner regions of the
disk.  This result is expected.  For Model B, the
mass to light ratio is $\simeq m^{1-y}_{{\scriptsize\rm u}}(m_{{\scriptsize\rm
u}}/m_{{\scriptsize\rm l}})^{-\Gamma -1}(y+\Gamma )(M_{\odot }/L_{\odot })/(-
\Gamma -1)$.  If
\s
\Gamma _{1}=[\ln (m_{{\scriptsize\rm u}}/m_{{\scriptsize\rm l}})\log _{10}{\it
e}]^{-1},
\labep{2}\e
\noindent which Model B obeys to within $15\%$, the mass to light ratio would
scale as
$\simeq e^{R/R_{0}}$.  This in turn would imply a disk surface density that is
relatively
constant.  The circular velocities of such disks increase monotonically with
$R$
and are zero at $R=0.$  This problem with low inner disk velocities is
probably
not serious because circular velocity curves are frequently
compatible even with constant mass to light ratio, halo-less models throughout
their entire optically-bright regions (e.g., Kent 1986).  Furthermore,
flatter, halo-less velocity curves could probably be attained by including the
following: galaxy
parameters slightly different than those of Kuijken \& Gilmore (1989a), a
(more
realistic$)
\log $-normal IMF (Miller \& Scalo 1979), expected spatial dependence in
remnant
and gas mass fractions, and variations of $m_{{\scriptsize\rm l}}$ or
$m_{{\scriptsize\rm u}}$ with $\rho _{{\scriptsize\rm l}}$.  For instance, the
velocity dip is better masked by the bulge if Bahcall \& Soniera's (1984)
smaller disk scale length of 3.5 kpc is assumed.
${\bf 5}{\hbox{\bf)}}$ The
IMF gradient of Model B appears to be too {\it small} to be compatible with
the
measurement of Garmany et al. (1982).  Equation ({}\ref{2}) suggests that this
discrepancy would be less if a smaller $m_{{\scriptsize\rm
u}}/m_{{\scriptsize\rm l}}$ ratio had been employed.
\par
\medskip
Of the above potential problems, only the first two appear to be
significant at this time.  Both can be overcome by taking into account the
arm/interarm density contrasts observed in spiral galaxies; Model C was
constructed to be similar to Model B, but has an azimuthally-averaged
light and mass density that is 3.25 times greater than the interarm values in
which the Sun presumably resides.  The circular velocity curve of Model C,
shown in the lower panel of Figure {}1, is slightly higher
but
otherwise similar to that of Model B. However, the solar-vicinity disk surface
density is only $60 M_{\odot }$pc$^{-2}$.  This is a much more reasonable 1.6
standard
deviations above the
value determined by Kuijken \& Gilmore (1989b) and is actually lower than
Bahcall \& Soniera's (1984) value of $\simeq 85 M_{\odot }$pc$^{-2}$.
\par
\noindent {}\setcounter{section}{4}\addtocounter{section}{-
1}\section{DISCUSSION}
\par
A direct scaling of R136's IMF to the Galaxy does not dramatically alter
the circular velocity curve.  However, Models B and C, with their higher yet
modest IMF gradients, have nearly flat $v_{\hbox{{\scriptsize\rm
circ}}}\lesssim 220$ km s$^{-1}$ circular velocity
curves only before the traditional dark halo component is included.  It is
interesting to note that if one assumes that these types of
models and their $\sim 10^{1}$-fold mass enhancements are representative of
most
galaxies, that the fiducial stellar contribution towards the closure
density is $\Omega _{*}\simeq 0.004~($e.g., Peebles 1993) before accounting
for IMF gradients,
that the cosmological constant is zero, and that there is no hot dark matter,
one would obtain $\Omega \simeq \Omega _{\hbox{{\scriptsize\rm baryon}}}\simeq
0.04+\Omega _{\hbox{{\scriptsize\rm gas}}}$, where the closure fraction due to
all
gas including hot plasma in galactic clusters is
$0.007\lesssim \Omega _{\hbox{{\scriptsize\rm gas}}}$$\lesssim 0.08~($Mulchaey
et al.
1996).
\par
\medskip
Current models of galactic evolution (e.g., Worthey 1994, de Jong 1996)
do not account for IMFs that might vary with time and position via the
temperature.  This is despite prior warnings that the IMF
probably has important dependences upon time and position (e.g., Mihalas \&
Binney 1978).  In light of the above results, accounting for IMFs with such
dependences may be necessary even to obtain results that are only accurate to
first order.  Accounting for these dependences may, for relatively obvious
reasons, clarify our understanding of several astrophysical phenomena
including the G-dwarf problem, intrinsic (as a function of radius) and
extrinsic (as a function of galactic morphology) metallicity and color
gradients, and the Tully-Fisher relation.
\par
\noindent \acknowledgments
\par
It is my pleasure to thank M. Harris, F. Bruhweiler, A. Whiting, A.
Fridman, D. Kazanas, D. Audley, J. Knapen, A. Mignogna, E. Albert, and A.
Stupp for useful comments.  This work is part of a dissertation to be
submitted to the Graduate School, University of Maryland in partial
fulfillment of the requirements of the Ph.D. degree in Physics.
\par

\par
\medskip
\inputfigurecaptionshere
\par
\noindent \newpage
\begin{table}
\begin{center}
\caption{IMFs in R136} 
\begin{tabular}{c c c c c} \tableline \tableline
$R$/pc & $\Gamma(R)$     & $m_{\rm l}$\qquad & $m_{\rm u}$ &
$\rho_{\rm l}/(L_\odot{\rm pc}^{-3})$ \\ \tableline
0.20 & $-1.29\pm 0.20$ & 5.6      & 120   & $1.5\times10^6$ \\
0.60 & $-1.46\pm 0.23$ & 3.6      & 76    & $1.5\times10^5$ \\
2.0 & $-2.12\pm 0.09$ & $\le2.0$ & 48 & $1.5\times10^3$ \\
\tableline
\end{tabular}
\tablecomments{Data adapted from Brandl et al. (1996) for (age-spread
restricted) stars 2.5---3.5 Myr old.}
\end{center}
\end{table}
\par
\medskip
\noindent \newpage
\begin{table}
\begin{center}
\caption{Model Parameters} 
\begin{tabular}{c c c c c c c c c c c} \tableline \tableline
Model & $\Gamma_0$ & $\Gamma_1$ & $m_{\rm l0}$ & $m_{\rm l1}$ & $m_{\rm u0}$ &
$m_{\rm u1}$ & $\rho_{\rm m\odot}/ $  & $\Sigma_{\rm
m\odot}/ $ & $<\Sigma_{\rm m\odot}>/
$ \\
&   &   &  &  &   &     & $M_\odot{\rm pc}^{-3} $  & $
M_\odot{\rm pc}^{-2} $ & $ M_\odot{\rm pc}^{-2} $ \\
\tableline
A & $-3.03 $ & $0.28 $ & $-0.08$ & $0.13$ & $1.25 $
& $0.12 $ & 0.12 & 67 & 67 \\
& $(\pm0.26)$ & $(\pm0.06)$ &  &  & $(\pm0.23)$ &
$(\pm0.04)$ \\  B & -1.11 & 0.42 & -1.52 & 0.00 & 1.30 & 0.00 & 0.31 & 179 &
179 \\
C & -0.55 & 0.40 & -1.52 & 0.00 & 1.60 & 0.00 & 0.10 & 60  &195
\\
\tableline
\end{tabular}
\end{center}
\tablecomments{This assumes $f=f_0+ f_1{\rm log}_{10}[\rho_{\rm
l}/(L_\odot{\rm
pc}^{-3})]$, for $f=\Gamma$, ${\rm log}_{10}(m_{\rm l})$, or ${\rm
log}_{10}(m_{\rm u})$.}
\end{table}
\par
\medskip
\noindent \inputfigureplotshere
\par
\noindent \end{document}
\end{document}